\documentclass[%
 reprint,
 amsmath,amssymb,
pra,
]{revtex4-1}

\usepackage{graphicx}% Include figure files
\usepackage{dcolumn}% Align table columns on decimal point
\usepackage{bm}% bold math
\graphicspath{%
    {converted_graphics/}% inserted by PCTeX
    {/}% inserted by PCTeX
}
\begin{document}

%\preprint{APS/123-QED}

\title{Fidelity of Physical Measurements}

\author{Thomas B. Bahder}
\affiliation{%
Aviation and Missile Research, 
      Development, and Engineering Center, \\    
US Army RDECOM, 
Redstone Arsenal, AL 35898, 
U.S.A.}%

\date{\today}

\begin{abstract}
The fidelity (Shannon mutual information between measurements and physical quantities) is proposed as a quantitative measure of the quality of physical measurements. The  fidelity  does not depend on the true value of  unknown physical quantities (as does the Fisher information) and  it allows for the role of prior information in the measurement process.  The fidelity is general enough to allow a natural comparison of the quality of classical and quantum measurements. As an example, the fidelity is used to compare the quality of measurements made by a classical and a quantum Mach-Zehnder interferometer.
\end{abstract}

%Use showkeys class option if keyword

\maketitle

%\tableofcontents

\section{\label{Intro}Introduction}

In any experiment, we want to determine the value of one or more physical quantities, say $x$, which can be one or more numbers.  However, in most experiments the actual quantities measured are not $x$ but some other quantities $y$.  We  infer the quantities $x$ from the measured quantities $y$, by using a  conditional probability, $P(y|x)$, that specifies the statistical relation between $x$ and $y$. The quantities $x$ and $y$ can be both discrete, both continuous, or any combination thereof.  The conditional probability, $P(y|x)$, gives the probability of a measurement \mbox{outcome $y$} given that the value of the physical quantity is $x$.  The probability $P(y|x)$ completely characterizes the measurement process, independent whether the system is classical or quantum.  In general, the conditional probability also  depends on one or more additional quantities, $\xi$, thereby having the form, $P(y|x, \xi)$, where the quantities $\xi$ label the state of the system at the time of measurement and the type of measurement that is performed.   

As an example, consider an experiment in which we want to measure the wavelength, $\lambda$,   of a light signal in units of nanometer.   It may happen that an experimenter has designed an apparatus to measure the wavelength $\lambda$, however the apparatus actually measures an electric current, $I$,  in units of ampere.  The conditional probability that describes this measurement apparatus is $P_1(I | \lambda)$.  Consider now a second experimentalist that designs another apparatus to measure the wavelength $\lambda$ using an alternate method.  In this alternate method, the apparatus may actually measure a voltage in units of volt.  This second apparatus is characterized by a conditional probability $P_2(V | \lambda)$, where $V$ is the measured voltage.  The question is: which apparatus is the best  for measuring the wavelength $\lambda$?  

In this letter, I propose to answer this question by comparing the fidelity (defined below) of each apparatus (measurement method).  The apparatus  with the highest value of fidelity provides, on average, the best measurement of $\lambda$.  The fidelity also takes into account our prior information about $\lambda$ through the prior probability distribution, $p(\lambda)$, see discussion below.

Historically, the quality of measurements of a quantity $x$ has been discussed in terms of parameter \mbox{estimation~\cite{Cramer1958,Helstrom1967,Helstrom1976,Holevo1982,Braunstein1994,Braunstein1996,Barndorff-Nielsen2000,Barndorff-Nielsen2003}}.  For example, for a single quantity $x$,  the Fisher information provides an upper bound on the variance, $\sigma^2_x$, of an unbiased estimator, $\hat{x}$, of the parameter $x$ through the Cram\'er-Rao inequality\cite{Cramer1958,Cover2006}
\begin{equation}
\sigma _x^2  \ge \frac{1}{{F_c(x)}}
\label{Cramer-Rao}
\end{equation}
where $F_c(x)$ is the classical Fisher information, defined by
\begin{equation}
F_{c}(x) = \sum\limits_y {\frac{1}{{P(y|x)}}\,\left[ {\frac{{\partial P(y|x)}}{{\partial x}}} \right]^2 } 
\label{FisherInformation}
\end{equation}
The quantity $F_{c}(x)$ depends on the type of measurement that is performed. For a quantum system, by maximizing over all possible measurement types,  Braunstein and Caves~\cite{Braunstein1994,Braunstein1996} have shown that a quantum Fisher information exists, such that $F_q(x) \ge F_c(x)$, and therefore an improved lower bound for $\sigma _x^2$ can be obtained by replacing $F_c(x)$ with $F_q(x)$ in Eq.~(\ref{Cramer-Rao}). The quantum Fisher information, $F_q(x)$, is defined by 
\begin{equation}
F_q \left( x \right) = {\rm tr} \left( \hat{\rho}_x \hat{\Lambda}_{x}^2  \right)
\label{QuantumFisherInfo}
\end{equation}
where $\hat{\Lambda}_{x}$ is the symmetric logarithmic derivative (SLD) that is implicitly given by~\cite{Helstrom1967,Helstrom1976,Holevo1982,Braunstein1994,Braunstein1996,Barndorff-Nielsen2000,Barndorff-Nielsen2003}.
\begin{equation}
\frac{{\partial \hat \rho _x }}{{\partial x}} = \frac{1}{2}\left[ {\hat \Lambda _x \,\hat \rho _x  + \hat \rho _x \hat \Lambda _x } \right]
\label{SLD}
\end{equation}

For a quantum measurement, the conditional probabilities can be obtained from 
\begin{equation}
P(y | x ) = {\rm tr} \left( \hat{\rho}_x \, \hat{\Pi}  \left( y \right) \right) 
\label{QuantumProbability}
\end{equation}
where the state is specified  by the density matrix, $\hat{\rho}_x$, and the measurements are defined by the positive-operator valued  measure (POVM),  $\hat{\Pi}  \left( y \right) $.   For classical measurements,  the conditional probabilities $P(y | x )$ can be obtained from a model of the classical measurement process, which may include phenomenological parameters characterizing noise in the measurements. 

The above description of the quality of measurements based on Fisher information is not satisfactory for two reasons.  First, the classical or quantum Fisher information, $F_c(x)$ or $F_q(x)$, may depend on the true value of the parameter $x$ if dissipation is present in the quantum system~\cite{Braunstein1996,Olivares2009,Gaiba2009,Bahder2010a}.  
Of course, the true value of the parameter $x$ is unknown. Second, the Fisher information does not take into account the prior information of the observer.  As an example, consider a child and an adult reading the same printed page of a book. Each of them may obtain a certain amount of information from the same printed page.  However, the  child  may obtain  less information from the printed page than the adult because the adult has more prior experience in the subject.  The Fisher information has no provision to take into account the observer's prior information.   

\section{\label{Fidelity}Shannon Mutual Information as Fidelity of Measurement}

I propose to use the fidelity as a quantitative measure of the quality of any physical experiment.  The fidelity is the Shannon mutual information~\cite{Shannon1948,Cover2006} between the measurements, $y$, and the physical quantities, $x$, \mbox{defined by} 
% \footnotesize
\begin{equation}
\small
\hspace{-0.02in} H[\xi] = \sum_y  \sum_x  P(y|x,\xi )   P(x)  \log_{2}\left[  
\frac{ P(y |x,\xi)}{\sum_{x^\prime}  P( y|x^{\prime},\xi )  P(x^\prime) } \right]
\label{ShannonMutualInformation}%
\end{equation}
where $P(y |x,\xi )$ is the conditional probability density of measuring $y$, given that the true value of the quantity is $x$, and given the state of the system and measurement type are specified by one or more quantities $\xi$.    
The fidelity gives the information (in bits) transferred between the quantity of interest, $x$, and the measurement result, $y$, for each use of the measurement apparatus.  The conditional probabilities  $P(y|x)$ can be obtained from a model (see below), or, from statistics of repeated experiments.  

Using the fidelity to characterize the quality of a measurement apparatus does not suffer from  the two objections to using the Fisher information, which were described above. The fidelity does not depend on the true value of the quantity $x$, because it is an average over all values of $x$ and $y$, using the conditional probabilities $P(y |x )$.  Furthermore, the fidelity depends on our prior knowledge about $x$ through the prior probability distribution  $P(x)$.  (If either $x$ or $y$, or both,  are continuous quantities, then the respective sums in Eq.~(\ref{ShannonMutualInformation}) are to be replaced by integrals.)   

The fidelity in Eq.(\ref{ShannonMutualInformation}) is a completely general way to characterize the quality of any classical or quantum measurement experiment. The classical or quantum measurement apparatus is a channel through which information flows from the phenomena, which is characterized by the value of the physical quantity $x$, to the measurements $y$.   

The fidelity gives the quality of the inference about the value of $x$ from the measurement of $y$.   However the fidelity  does not give an estimate of the value of $x$.  The value of the quantity $x$ can be inferred from the probability distribution for $x$, using Bayes' rule 
\begin{equation}
P(x|y, \xi) = \frac{{P(y|x, \xi)P(x, \xi)}}{{\sum\limits_x {P(y|x,\xi)P(x,\xi)} }}
\label{BayesRule}
\end{equation}
where I have included the dependence on other parameters $\xi$.  The  value of the quantity $x$ can be estimated, for example, by taking the mean of the distribution given by Eq.~(\ref{BayesRule}).  Once we have made an estimate of the value of $x$,  we can improve on this estimate by making recursive measurements. The distribution for the value of $x$, $P(x|y, \xi)$ given in  Eq.~(\ref{BayesRule}),  can be used as our new prior probability distribution, setting $P(x)=P(x|y,\xi)$  in  Eq.(\ref{ShannonMutualInformation}).  Furthermore, the fidelity can be maximized with respect to $\xi$  for the next measurement, using our current state of knowledge, represented by $P(x|y,\xi)$.  In this way, we can optimize a measurement device (classical or quantum) to give the best possible measurement in the next measurement cycle.    

The fidelity has already been used to discuss the quality of phase determination in quantum interferometers~\cite{Bahder2006,Bahder2007,PhysRevA.78.053829,Bahder2010a} and to discuss the sensitivity to rotation of classical Sagnac gyroscopes~\cite{Bahder2010b}.  

\section{\label{FidelityComparison}Comparison of Classical and Quantum Measurements}

The fidelity can be used to determine which experiment (apparatus) provides a better measurement of a given physical quantity. As an example, I compare the fidelity of a quantum and a classical measuring device.   Specifically, I compare a classical and a quantum Mach-Zehnder (\mbox{M-Z}) interferometer, each of which can be used to determine the phase $\phi$ in one arm of the interferometer. For the classical \mbox{M-Z} interferometer, the direct measurement is the energy in each of the output ports, $E_c$ and $E_d$, which are continous variables, see discussion below.  For a quantum \mbox{M-Z} interferometer, the direct measurement is the integer number of photons in the output ports, $n_c$ and $n_d$.   So in the notation above, the phase $\phi$ plays the role of $x$ and the measurements $y$ are the pair of numbers, $(E_c, E_d)$, see  Eq.~(\ref{ShannonMutualInformation}). 

\hspace{-1.0in} Consider a quantum Mach-Zehnder interferometer with input ports labeled ``a" and ``b" and output ports labeled ``c" and ``d". Assume we input the state 
\begin{equation}
 |\alpha \rangle _a\otimes |0\rangle _b  =e^{-\frac{1}{2}|\alpha |^2}\sum _{n=0}^{\infty } \frac{\alpha ^n}{(n!)^{1/2}} |n\rangle_a\otimes |0\rangle _b 
\label{coherentState}
\end{equation}
which consists of  a coherent state input into port ``a" and vacuum input into port ``b", where $a$ and $b$ label the modes,  $| n \rangle$ is a Fock state of $n$ photons, and the complex parameter $\alpha $ specifies the average photon number and photon number variance,  $|\alpha|^2 = \langle \hat{n} \rangle = \langle (\Delta \hat{n})^2 \rangle $.   The probability that $n_c$ and $n_d$ photons are output in ports ``c" and ``d", respectively, is~\cite{Bahder2006}
\begin{equation}
\small
 P\left( \left. n_{c},n_d\right|\phi ,\alpha \right)=\frac{e^{-|\alpha |^2}}{n_c!n_d!}
|\alpha |^{2\left(n_c+n_d\right)}\sin ^{2 n_c}\left(\frac{\phi }{2}\right)\cos ^{2 n_c}\left(\frac{\phi }{2}\right)
\label{ProbCoherentPhotonOutput}
\end{equation}
where $\phi$ is the phase shift in one arm of the interferometer.  The average  energy of this coherent state is $ \langle E\rangle = \langle n\rangle \hbar \omega  = \hbar \omega |\alpha |^2 $  with energy spread \mbox{$\Delta E = \hbar \omega | \alpha|=\sqrt{\hbar \omega} \, \langle E \rangle ^{1/2} $}. The discrete energies output in port ``c" and ``d" are $E_c = n_c \hbar \omega$ and $E_d = n_d \hbar \omega$, respectively.  The joint probability density  for measuring energy $E_c$ and $E_d$ output in ports ``c" and ``d", respectively, is given by 
\begin{equation}
p(E_c,E_d | \phi, \bar{E}) = \frac{P\left(\left.n_{c,}n_d\right|\phi ,\alpha \right)}{ ( \hbar \omega )^2}
\label{ProbEnergyOutputInCoherentStateInput}
\end{equation}
where       $P\left(\left.n_{c,}n_d\right|\phi ,\alpha \right)$    is given in Eq.~(\ref{ProbCoherentPhotonOutput}) and  I use the notation for the average energy \mbox{$\bar{E} = \langle E \rangle$}. Using Eq.~(\ref{ProbCoherentPhotonOutput}) in Eq.~(\ref{ShannonMutualInformation}), and assuming no prior knowledge about phase, taking $p(\phi)=1/(2 \pi)$, I find the fidelity for the quantum \mbox{M-Z} interferometer with coherent state input to be: 
\begin{widetext}
\small
\begin{equation}
H_{\rm coh}(|\alpha|^2)  =   \frac{e^{-|\alpha |^2}}{2\pi }\sum _{n_c=0}^{\infty } \sum _{n_d=0}^{\infty } \frac{|\alpha |^{2\left(n_c+n_d\right)}}{n_c! n_d!}\int _{-\pi }^{+\pi } d\phi \, \sin ^{2 n_c}\left(\frac{\phi }{2}\right)\cos ^{2 n_d}\left(\frac{\phi }{2}\right)  
   \log _2\left[ \frac{\pi \left(n_c+n_d\right)!}{\Gamma \left(n_c+\frac{1}{2}\right)\Gamma \left(n_d+\frac{1}{2}\right)}\sin ^{2 n_c}\left(\frac{\phi }{2}\right)\cos ^{2 n_d}\left(\frac{\phi }{2}\right)\right] 
\label{FidelityQuantumMZ}
\end{equation}
\end{widetext}
which is only a function of the parameter $|\alpha|^2 $.

Now consider a classical  \mbox{M-Z} interferometer with a finite-duration pulse of monochromatic light of energy $E_n$ input into port ``a" and vacuum input into port ``b". I assume that the pulse has a duration in time sufficiently long that I can describe the pulse as  having a single frequency and  energy $E_n$.  We want to determine the phase $\phi$, however, the direct measurement consists of the energies in the output ports, $E_c$ and $E_d$.    The classical \mbox{M-Z} interferometer is defined by  energies $E_c$ and $E_d$ output in ports ``c" and ``d", respectively
% \begin{eqnarray}  %%%% XXXXXXXXXXXXXXXXXXXXXXXXXXXXXXXXXXX
% E_c  & = &  E_n \sin^2 \left(\frac{\phi}{2} \right) \nonumber \\
% E_d & =  &  E_n \cos^2 \left(\frac{\phi}{2} \right)  \label{ClassicalMZOutputB}
% \end{eqnarray}
%
\begin{equation}
E_c   =   E_n \sin^2 \left(\frac{\phi}{2} \right), \hspace{0.4in} E_d  =    E_n \cos^2 \left(\frac{\phi}{2} \right)  
\label{ClassicalMZOutputB}
\end{equation}
where $\phi$ is the phase in one arm of the \mbox{M-Z} interferometer.
In order to compute the fidelity of the classical \mbox{M-Z} interferometer, we need to define a classical measurement model.  In the quantum interferometer described above in Eqs.~(\ref{coherentState})---(\ref{FidelityQuantumMZ}), the input state had a spread in energy $\Delta E$ due to photon number fluctuations. For the case of the classical \mbox{M-Z} interferometer, I assume a probability distribution, $P_a\left(E_n\right)$, of closely-spaced discrete input energies,  $E_n = n \, \delta E$, where $n=0,1,2,\cdots, N_E$, where $N_E$ is the number of energies in the energy grid, and $\delta E $ is an arbitrary discrete energy step, which can be taken to zero later.  I also take the phase as a discrete set of $ 2 N_\phi$  possible values, \mbox{$\phi_k = \pi k /N_\phi$, for $k= \{ 0,\pm 1, \pm2, \cdots, \pm N_\phi -1, N_\phi \}$}, where $\Delta \phi = \phi_{k+1} - \phi_k$ is the uniform grid spacing of phase.  Therefore, I take the conditional probability for a classical measurement outcome, $(E_c,E_d)$, to be
\begin{equation}
 P\left(E_c,\left.E_d\right|\phi_k \right) =\sum _{E_n} P\left(E_c,\left.E_d\right|\phi_k ,E_n\right) \, P_a\left(E_n\right)
\label{ProbabilityClassicalMeasurement}
\end{equation}
where  $E_c = E_c(n,k) = n \, \delta E \sin^2(\phi_k/2)$ and $E_d = E_d(n,k) = n \, \delta E \cos^2(\phi_k/2) $, are discrete energies measured in output ports ``c" and ``d", respectively, given the energy $E_n = n \, \delta E$ is input in port ``a"  and vacuum is input in port ``b".  Equation~(\ref{ProbabilityClassicalMeasurement}) is a special case of the identity between conditional probabilities, $P(B|C) = \sum_A  P(B|A,C) \, P(A|C)$.  Note that the discrete energies  $E_c(n,k)$ and $E_d(n,k)$ depend on a two indices, $n$ and $k$, each of which have different ranges, depending on the energy grid and the phase grid.   The probability distribution for input energies is normalized, $\sum_{n=0}^{\infty} P_a\left(E_n\right) =1$.    The conditional probability of measuring discrete energies  $E_c$ and $E_d$ in the output ports, given monochromatic input energy $E_m = m \, \delta E$ and phase $\phi _l$, can be written as a product of Kr\"onecker delta functions:
\begin{equation}
P\left(E_c(n,k),E_d(n^\prime,k^\prime)|\, \phi _l,E_m\right)= \delta_{n,m} \, \delta_{k,l}, \, \delta_{n^\prime,m} \, \delta_{k^\prime,l}
\label{ProbabilityClassicalMeasurementMonochromatic}
\end{equation}
where where $n,n^\prime, m = 0,1,2,\cdots, N_E$ and  $k,k^\prime,l= \{ 0,\pm 1, \pm2, \cdots, \pm N_\phi -1, N_\phi \}$.  Using Bayes' rule in Eq.~(\ref{BayesRule}), the phase probability distribution is given by 
\begin{equation}
P\left(\phi _l|E_c(n,k),E_d(n,k)\right)=\delta _{k,l}
\label{PhaseDistributionDiscrete}
\end{equation}
In the limit $\Delta \phi \to 0$ of a continuous phase variable, the phase probability density, $p\left(\phi \left|E_c\right.,E_d\right) \equiv 
 \left.P\left(\phi _l|E_c(n,k),E_d(n,k)\right)\right/\Delta \phi $, is given by the right side of Eq.~(\ref{phaseProbDenisty}), which gives two values of phase for each classical measurement outcome  $(E_c, E_d)$. 

Using Eq.~(\ref{ProbabilityClassicalMeasurement}) and (\ref{ProbabilityClassicalMeasurementMonochromatic}) in Eq.~(\ref{ShannonMutualInformation}), I find the fidelity of this classical \mbox{M-Z}  interferometer to be
\begin{equation}
H=  \frac{2\pi }{\Delta \phi }\log _2\left(\frac{2\pi }{\Delta \phi }\right)
\label{FidelityClassicalDiscretePhi}
\end{equation}
where $2\pi / \Delta \phi$ is the number of phase points in the range $-\pi < \phi \le + \pi$.   The fidelity in Eq.~(\ref{FidelityClassicalDiscretePhi}) is independent of the input energy  probability distribution $P_a\left(E_n\right)$.   As $\Delta \phi \rightarrow 0 $, the fidelity in Eq.~(\ref{FidelityClassicalDiscretePhi}) diverges  because there are no fluctuations or energy measurement errors built into the classical measurement model in Eq.~(\ref{ProbabilityClassicalMeasurement}) and (\ref{ProbabilityClassicalMeasurementMonochromatic}). This classical measurement model assumes that energy measurements are arbitrarily precise.  In reality, there is noise in energy measurements that limits the phase resolution, leading to a non-zero  value $\Delta \phi$ that makes the fidelity $H$ finite.   

An  improvement over the classical measurement model in Eq.~(\ref{ProbabilityClassicalMeasurement}) and (\ref{ProbabilityClassicalMeasurementMonochromatic}) can be made by assuming that the  probability of classical energy measurement is not sharp but instead has some statistical error of order $\Delta$ due to unmodelled physical processes. The value of the phenomenological parameter $\Delta$ can be obtained from experiments by taking $\Delta$ equal to the standard deviation of classical energy measurements in the \mbox{M-Z} interferometer.  As an improved classical measurement model, I take  the probability density for measuring energy $E_c$ and $E_d$ in output ports ``c"  and ``d" respectively, as
\begin{figure}[tbp] % float placement: (h)ere, page (t)op, page (b)ottom, other (p)age
  \centering
  % file name: C:/Documents and Settings/BahderTB/My Documents/cherry/Fidelity_of_Measurements/Hplots.eps
   \includegraphics[width=3.3in]{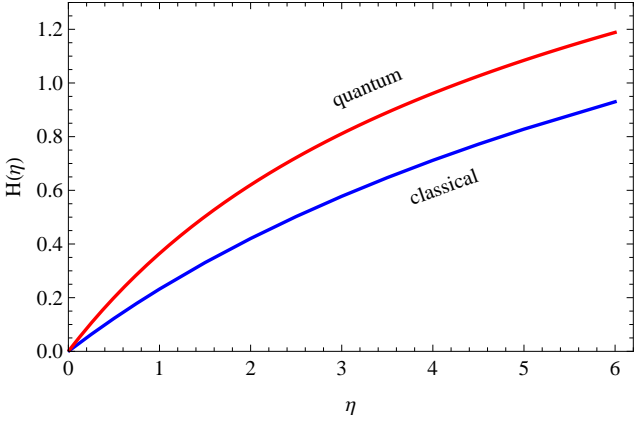}
  \caption{The fidelity is plotted (in units of bits) for quantum (red line) and classical (blue line) interferometers, $H_{\rm coh}(\eta)$ and $H_{\rm class}(\eta)$, respectively, vs. $\eta$, where $\eta$ is the dimensionless energy in units of photon number. No prior knowledge of phase was assumed, taking $p(\phi)=1/(2 \pi)$. The quantum \mbox{M-Z} interferometer has a higher fidelity than the classical \mbox{M-Z} interferometer, showing that the quantum interferometer gives more information on phase $\phi$ than the classical interferometer. 
  \label{fig:Hplots}}
\end{figure}
\begin{eqnarray}
p\left(\left.E_c\right|\phi ,E\right) & = & \int _{-\infty }^{+\infty }p\left(\left.E_c\right|\phi, E ,\varepsilon _c\right)p_{\Delta }\left(\varepsilon _c\right)d\varepsilon _c \nonumber \\
p\left(\left.E_d\right|\phi ,E\right) & = & \int _{-\infty }^{+\infty }p\left(\left.E_d\right|\phi, E ,\varepsilon _d\right)p_{\Delta }\left(\varepsilon _d\right)d\varepsilon _d  
\label{ClassicalMeasurementModelForC}
\end{eqnarray}
where $p\left(\left.E_c\right|\phi, E ,\varepsilon _c\right)$ is the conditional probability that energy $E_c$ is measured in port ``c", given the phase $\phi$, the  input energy $E$ and the error in the classical measurement $\varepsilon_c$, with analogous definitions for port ``d".   Here, $ p_{\Delta }(\varepsilon )$ is a normal probability distribution that introduces errors into the measurement of energies $E_c$ and $E_d$: 
\begin{equation}
 p_{\Delta }(\varepsilon )=\frac{1}{\sqrt{2\pi } \Delta } \exp \left({-\frac{\varepsilon ^2}{2\Delta ^2}} \right)
\label{ClassicalErrorModel}
\end{equation}
Equation~(\ref{ClassicalMeasurementModelForC}) is the result of an identity for conditional probabilities.  In view of the classical output relations in Eq.~(\ref{ClassicalMZOutputB}), I take the conditional probability density for  measuring the output energies $E_c$ and $E_d$  to be defined in terms of Dirac $\delta$ functions:
\begin{eqnarray}
p\left(\left.E_c\right|\phi ,E, \varepsilon_c\right) & = & \delta(E_c - E \sin^2\left( \frac{\phi}{2}\right) -\varepsilon_c  ) \nonumber \\ 
p\left(\left.E_d\right|\phi ,E,\varepsilon_d\right) & = & \delta(E_d - E \cos^2\left( \frac{\phi}{2}\right) -\varepsilon_d  ) 
\label{monInputClassical}
\end{eqnarray}
Taking  the product of the distributions in Eq.~(\ref{ClassicalMeasurementModelForC}) leads to the conditional joint probability density for a classical measurement outcome, $(E_c,E_d)$, given the phase is $\phi$ and monochromatic energy $E$ is input: 
\begin{eqnarray}
 & & p\left(E_c,\left.E_d\right|\phi ,E,\Delta \right)  =  p\left(\left.E_c\right|\phi ,E\right) \, p\left(\left.E_d\right|\phi ,E\right)  \nonumber \\
\hspace{-0.4in}  = &  & p_{\Delta }\left(E_c-E \sin ^2\left(\frac{\phi }{2}\right)\right) p_{\Delta }\left(E_d-E \cos ^2\left(\frac{\phi }{2}\right)\right)
\label{classicalMeasurementOutcomeProbabiliyDensity}
\end{eqnarray}
In Eq.~(\ref{classicalMeasurementOutcomeProbabiliyDensity}), the energies $E_c$ and $E_d$ are continuous variables, as is the phase $\phi$.  From Bayes' rule in Eq.~(\ref{BayesRule}), assuming no prior information on phase, therefore taking $p(\phi)=1/2 \pi$, the phase probability density is
\begin{equation}
 p\left(\phi \left|E_c\right.,E_d,E,\Delta \right)=\frac{p\left(E_c,\left.E_d\right| \phi ,E,\Delta \right)}{\int_{-\pi }^{+\pi } p\left(E_c,\left.E_d\right| \phi ,E,\Delta \right) \, d\phi }
\label{phaseProbDenisty}
\end{equation}
The phase probability density, $p\left(\phi \left|E_c\right.,E_d,E,\Delta \right)$, has two peaks, and as $\Delta \rightarrow 0$ it approaches the sum of two $\delta$ functions:
\begin{eqnarray}
 p\left( \phi \left| E_c  \right. , E_d , E, \Delta  \right)  & \to  & \frac{1}{2} \left[ \,\, \delta \left( \phi  - 2\arctan \sqrt{  \frac{E_c}{E_d} } \right)   \right. \nonumber \\ 
   &   & \hspace{-0.5in}  \left. + \delta \left( \phi  + 2\arctan \sqrt{ \frac{E_c}{E_d} } \right) \,\, \right] 
 \label{phaseProbDenisty2}
\end{eqnarray}
%
%\begin{eqnarray}
% p\left( {\phi \left| {E_c } \right.,E_d ,E,\Delta } \right)  & \to  & \frac{1}{2}\left[ {\delta \left( \phi  - 2\arctan \left( \left( \frac{{E_c }{E_d}}  \right)} \right)  } \right. \nonumber \\ 
%   &   & \hspace{-0.5in}  \left. + {\delta \left( {\phi  + 2\arctan \left( {\left( {\frac{{E_c }}{{E_d }}} \right)^{1/2} } \right)} \right)} \right] 
% \label{phaseProbDenisty}
%\end{eqnarray}
%
Trivially changing the sums to integrals in the definition of fidelity in Eq.~(\ref{ShannonMutualInformation}), and using Eq.~(\ref{classicalMeasurementOutcomeProbabiliyDensity}), leads to the fidelity for a classical \mbox{M-Z} interferometer:

%
%\begin{equation}
% H_{\rm class}(E,\Delta) = \int _{-\infty }^{+\infty }dE_c\int _{-\infty }^{+\infty }dE_d \int _{-\pi }^{+\pi }d\phi \,\,
% p\left(E_c,\left.E_d\right|\phi ,E,\Delta \right) p(\phi ) \log _2\left(\frac{p\left(E_c,\left.E_d  \right|\phi ,E, \Delta \right)}  {p\left(E_c,E_d, E, \Delta\right)}  \right) \,\,\,\,
% \label{ClassicalFidelityContinuousVars222222222}
% \end{equation}
%
%
%
%
\begin{eqnarray}
 & & H_{\rm class}(E,\Delta) = \int _{-\infty }^{+\infty }dE_c\int _{-\infty }^{+\infty }dE_d \int _{-\pi }^{+\pi }d\phi  \times \nonumber \\
  &  & p\left(E_c,\left.E_d\right|\phi ,E,\Delta \right) p(\phi ) \log _2\left(\frac{p\left(E_c,\left.E_d  \right|\phi ,E, \Delta \right)}  {p\left(E_c,E_d, E, \Delta\right)}  \right) \,\,\,\,
\label{ClassicalFidelityContinuousVars}
\end{eqnarray}
where
\begin{equation}
\hspace{-0.015in} p\left(E_c,E_d, E, \Delta \right)= \int _{-\pi }^{+\pi }p\left(E_c,\left.E_d\right|\phi ,E, \Delta \right) p(\phi ) d\phi 
\label{ClassicalFidelityContinuousVarsNormalizationFunction}
\end{equation}
and $p\left(E_c,\left.E_d\right|\phi ,E, \Delta \right)$ is given by Eq.~(\ref{classicalMeasurementOutcomeProbabiliyDensity}), and $p(\phi)$ is the probability representing our prior knowledge about~$\phi$.
Equation~(\ref{ClassicalFidelityContinuousVars}) gives the fidelity of a classical \mbox{M-Z} interferometer with monochromatic input energy $E$ and errors in energy measurements of order $\Delta$. The errors, $\varepsilon_c$ and $\varepsilon_d$, in energy measurements, $E_c$ and $E_d$, can be imagined as due to unmodelled noise (e.g., shot noise) in the measurements.  As $\Delta \rightarrow 0$, the measurements have no noise (errors)  and the fidelity $H_{\rm class}(E,\Delta) \rightarrow \infty $, compare with Eq.~(\ref{FidelityClassicalDiscretePhi}) for the case $\Delta \phi \to 0$.

The fidelity for the classical interferometer, $H_{\rm   class}(E,\Delta)$ in Eq.~(\ref{ClassicalFidelityContinuousVars}), depends on two parameters, $E$ and $\Delta$, and on our knowledge of $\phi$ given by the prior phase distribution, $p(\phi)$. The fidelity for the quantum interferometer, $H_{\rm coh}(|\alpha|^2)$ in Eq.~(\ref{FidelityQuantumMZ}), depends on only one parameter, $\eta$, where we assumed no prior knowledge about phase by taking $p(\phi)=1/(2 \pi)$.     A direct comparison of the fidelity of the quantum and classical \mbox{M-Z} interferometers can be made by assuming in the classical case in Eq.~(\ref{ClassicalFidelityContinuousVars}) no prior knowledge of the phase taking $p(\phi)=1/(2 \pi)$, and taking the energy  $E=\hbar \omega |\alpha|^2 \equiv \hbar \omega \eta$ and energy width  $\Delta = \sqrt{\hbar \omega E} \equiv \hbar \omega \sqrt{\eta}$, which gives the same energy width for the measurements of the classical \mbox{M-Z} interferometer as for the coherent input state of the quantum \mbox{M-Z} interferometer.   For the classical \mbox{M-Z} interferometer, $\eta$ is the monochromatic input energy in units of photon energy, $\hbar \omega$. For the quantum interferometer, $\eta$ is the average energy $\langle E \rangle $ of the input coherent state in units of photon energy, $\hbar \omega$.    With this parametrization, the  fidelity of the classical \mbox{\mbox{M-Z}} interferometer,  $H_{\rm class}(\hbar \omega \eta, \hbar \omega \sqrt{\eta})$, depends only on $\eta$, and can be directly compared to the fidelity of the quantum interferometer, $H_{\rm coh}(\eta)$, see Fig.~1.  It is clear that, for a single use of the interferometer, the quantum measurement  (apparatus) has a higher fidelity (provides more bits of information about the phase) than the classical measurement. 

\section{\label{Conclusion}Conclusion}

Two objections have been raised against using the Fisher information as a measure of the quality of measurements.  First, the Fisher information may depend  on the unknown physical quantity (parameter to be determined), which may occur when dissipation is present.  Second, the Fisher information does not take into account  prior information about the parameter.  Consequently, I  proposed the use of  fidelity (Shannon mutual information between measurements and physical quantities)  in Eq.~(\ref{ShannonMutualInformation})  as a quantitative measure of the quality of physical measurements.   The fidelity does not depend on the value of the unknown physical quantity because it is an average over all probability distributions of that quantity.   Also, the fidelity takes into account an observer's prior information through the prior probability distribution,  $P(x)$, see Eq.~(\ref{ShannonMutualInformation}).  The dependence on prior information also allows us to update recursively our information during repeated  experiments.  Also, the fidelity can be maximized with respect to (classical or quantum) measurements, parameters in the experiment, and with respect to (classical or quantum) input states.  Finally, the fidelity is general enough to quantitatively compare the quality of classical and quantum measurements, or to compare two different experiments that attempt to determine the same physical quantity.   As an example of this, I have considered a quantum \mbox{M-Z} interferometer with a coherent state input into one port, and I have compared it with a classical interferometer with phenomenological error in measuring the energy in the output ports. For the range of parameters considered, see Fig.~\ref{fig:Hplots},  the quantum \mbox{M-Z} interferometer has higher fidelity than the classical interferometer,  indicating that, for each measurement the quantum \mbox{M-Z} interferometer provides more bits of information on the phase than the classical \mbox{M-Z} interferometer.  The fidelity allows a quantitative comparison of the quality of these two types of measurements.    Finally, non-ideal aspects of experiments, such as non-deterministic state creation,   absorption, and errors in measurements (e.g., photon counting errors or energy measurement errors) can be included in the fidelity by using the formalism that was developed in Ref.~\cite{Bahder2010a}.

% \begin{acknowledgments}
%  xxxx
%  \end{acknowledgments}
%
%%%%%%%%%%%%%%% REFERENCES  %%%%%%%%%%%%%%%%%%%%
\bibliographystyle{apsrev}
\bibliography{References-Quantum}
%%%%%%%%%%%%%%%%%%%%%%%%%%%%%%%%%%%%%%%%%%%%%%%%
%
\end{document}